\documentclass[preprint,onecolumn,nofootinbib]{revtex4}
\pdfoutput=1
\usepackage[colorlinks=true,linkcolor=blue,urlcolor=blue,filecolor=black,citecolor=red,pdfstartview=FitV,pdftitle={},pdfsubject={},pdfkeywords={},pdfpagemode=None,bookmarksopen=true]{hyperref}
\usepackage{graphicx}
\usepackage{amsmath}
\usepackage{amsfonts}
\usepackage{amssymb,ulem}
\usepackage{color,xcolor}%
\usepackage{CJK}
\usepackage{subfigure}

\usepackage{dcolumn}
\newcommand{\f}{\begin{equation}}
\newcommand{\ff}{\end{equation}}
\newcommand{\fa}{\begin{eqnarray}}
\newcommand{\ffa}{\end{eqnarray}}

\begin{document}
\title{Holographic phonons by gauge-axion coupling}

\author{Xi-Jing Wang$^{1}$}
\thanks{xijingwang@yzu.edu.cn}
\author{Wei-Jia Li$^{2}$}
\thanks{weijiali@dlut.edu.cn} \affiliation{
  $^1$ Center for Gravitation and Cosmology, College of Physical
  Science and Technology, Yangzhou University, Yangzhou 225009,
  China\\
  $^2$ Institute of Theoretical Physics, School of Physics, Dalian University of Technology, Dalian 116024, China.
}

\begin{abstract}
In this paper, we show that a simple generalization of the holographic axion model can realize spontaneous breaking of translational symmetry by considering a special gauge-axion higher derivative term. The finite real part and imaginary part of the stress tensor imply that the dual boundary system is a viscoelastic solid. By calculating quasi-normal modes and making a comparison with predictions from the elasticity theory, we verify the existence of phonons and pseudo-phonons, where the latter is realized by introducing a weak explicit breaking of translational symmetry, in the transverse channel.  Finally, we discuss how the phonon dynamics affects the charge transport.
\end{abstract}
\maketitle
\tableofcontents

\section{Introduction}
It is widely unveiled that the hydrodynamic formulation, which is based upon a small amount of conserved quantities as well as the constitutive relations, can provide a universal description for the low energy collective excitations in exotic quantum matter \cite{Teaney:2009qa,fermigas,graphene1,graphene2}. In relativistic hydrodynamics \cite{Kovtun:2012rj}, the most fundamental conserved quantity is the stress tensor. Despite the Poincar\'e symmetry is  fundamental for particles in the high energy region, the spatial translations are ubiquitously broken at low energy scales in condensed matter physics, especially in solids where there commonly exist periodic lattice, impurities, defects, etc. If the breaking of translations is weak, we are still allowed to formulate the so-called quasi-hydrodynamics to address the problems approximately. However, this framework might break down when the translational symmetry is strongly violated.

On the contrary, the holographic duality provides a purely geometric approach to understand universal behaviors in strongly coupled systems, regardless of whether quantities are conserved  or not \cite{Hartnoll:2009sz,Natsuume:2014sfa,McGreevy:2016myw,Hartnoll:2016apf}. In holography, all the physics of a many-body system (which is described by some quantum field theories) is encoded in a dual gravity theory with one extra dimension. Moreover, in the strong coupling limit of the field theory, the quantum effects on the gravity side are significantly suppressed. As a result, complicated problems on the field theory side are simplified into solving classical equations in a curved spacetime. Consequently, this duality has nowadays become a widely used tool for exploring strongly coupled many-body systems.

Despite the earlier holographic studies focused more on translation invariant systems, some recent work attempts to make setups closer to realistic materials without translations \cite{Horowitz:2012ky,Horowitz:2012gs,Donos:2012js,Vegh:2013sk,Horowitz:2013jaa,Blake:2013bqa,Ling:2013nxa,Donos:2013eha,Andrade:2013gsa,Donos:2014uba,Gouteraux:2014hca,Ling:2014laa,Davison:2014lua,Donos:2019tmo}. One of the simplest ways to capture the key features of the symmetry breaking patterns of translations but allow fast calculations is to consider homogeneous massive gravity models. Based on the perturbative calculations, it has also been explicitly shown that gravitons always have a non-zero mass in any holographic system with an inhomogeneous lattice \cite{Blake:2013owa}. For this reason, massive gravity provides a universal low energy description for strongly coupled systems with broken translations. A generally and widely used way of modeling massive gravity is to consider the axions in the bulk \cite{Baggioli:2014roa,Alberte:2014bua,Alberte:2015isw,Baggioli:2016oju,Alberte:2017cch}. These are massless scalars enjoying a internal shift symmetry, linearly depending on spatial coordinates. As a result, the background configuration of the axions contributes a non-zero mass to gravitons, breaking the translations in the boundary system but retaining the homogeneity of the background. Better yet, one can easily realize explicit breaking (ESB) or spontaneous breaking (SSB) of translations differently by slightly changing the kinetic term of the axions in the bulk \cite{Alberte:2017oqx,Ammon:2019wci,Ammon:2019apj,Baggioli:2019abx}  \footnote{If the breaking of translations is purely spontaneous, there should exist gapless excitations in the low energy description called Goldstone modes, also known as phonons in solid. It is worth studying the situation where the breaking of translations is mostly but not totally spontaneous, which is called pseudo-spontaneous breaking of translations. The corresponding excitations are so-called pseudo-Goldstone modes or pinned phonons.}. For a comprehensive review on the holographic axion models, one refers to \cite{Baggioli00:2021xuv}.

In this work, we consider a new way of spontaneously breaking translations by directly coupling the axions to the $U(1)$ gauge field at finite density. Note that such types of gauge-axion higher derivative terms have previously introduced in \cite{Gouteraux:2016wxj,Baggioli:2016pia} in the holographic studies on the metal-insulator transition (MIT). Here, we consider the simplest one of them and reveal the fact that this term just introduces a spontaneous breaking of translations and the pinning effect in presence of external sources leads to the MIT. Moreover, different from the fluid model discussed in \cite{Li:2018vrz}, the gravity system under consideration is dual to some solid states on boundary where there exist propagating/pinned modes in the transverse channel. The paper is organized as follows: In Section \ref{bgsolution}, we introduce the holographic setup, solve the background solution and analyze the viscoelasticity of the dual system. In Section \ref{lespectrum}, the spectrum of transverse phonons is systematically investigated by computing quasi-normal modes (QNMs) of the black hole. In Section \ref{ctransport}, we study how the phononic dynamics impacts the charge transport. In Section \ref{conclusion}, we conclude.

\section{An elastic black hole}\label{bgsolution}
\subsection{Spontaneous breaking of translations}
In this paper, we consider the following action which can break translational symmetry of dual field theories \citep{Gouteraux:2016wxj}
\begin{equation}
S=\int d^4 x \sqrt{-g}\left[R-2\Lambda-\lambda X-\frac{1}{4}\left(1+\mathcal{K}X\right)F^2\right],\label{action}
\end{equation}
where $X$ is the kinetic term of the axions $\phi^I$ and $F_{\mu\nu}$ is $U(1)$ gauge field which are described by
\begin{equation}
X=\frac{1}{2}\sum_{I=x,y}\partial^{\mu}\phi^I\partial_{\mu}\phi^I,  \qquad  F_{\mu\nu}=\nabla_{\mu}A_{\nu}-\nabla_{\nu}A_{\mu}.
\end{equation}
And, $\Lambda$ is the cosmological constant that will be set  $\Lambda=-3$ for simplicity. $\lambda$ and $\mathcal{K}$ are dimensionless couplings. If we set $\mathcal{K}=0$, this action simply reduces to the simple \textit{linear axion model} \cite{Andrade:2013gsa}. In our model, we should require that $\lambda \geq 0$ for avoiding the ghost problem and $-1/6\leq \mathcal{K}\leq 0$ for causality and stability \cite{Baggioli:2016oqk,Gouteraux:2016wxj} \footnote{In the paper \cite{Gouteraux:2016wxj}, the author obtained the coupling $-1/6 \leq  \mathcal{K} \leq 1/6$. While in \cite{Baggioli:2016oqk}, the author gave the stronger constraint $-1/6 \leq  \mathcal{K} \leq 0 $ to avoid the problem of ghosts. In these discussions the effect of magnetic field is not included. As was pointed in \cite{An:2020tkn} that turning on an external magnetic field will make the constraint on $\mathcal{K}$ even tighter. However, in this paper, our construction does not involve the magnetism.}.

Varying the bulk fields, their equations of motion are given by
\begin{gather}
R_{\mu\nu}-\frac{1}{2}g_{\mu\nu}R-\frac{1}{2}(\lambda+\frac{\mathcal{K}}{4}F_{\sigma\rho}F^{\sigma\rho})\nabla_\mu \phi^I\nabla_\nu \phi^I-\frac{1}{2}\left(6-\frac{\lambda}{2}\nabla_\sigma \phi^I \nabla^\sigma \phi^I\right)g_{\mu\nu}\\ \nonumber
=\frac{1}{2}\left(1+\frac{\mathcal{K}}{2}\nabla_\sigma \phi^I \nabla^\sigma \phi^I\right)\Big({F_\mu}^{\sigma}F_{\nu\sigma}-\frac{1}{4}g_{\mu\nu}F_{\rho\sigma}F^{\rho\sigma}\Big),\\ \nonumber
\\ 
\nabla_\mu\left[\left(1+\frac{\mathcal{K}}{2}\nabla_\sigma \phi^I \nabla^\sigma \phi^I\right)F^{\mu\nu}\right]=0,
\end{gather}
and
\begin{gather}
\nabla_\mu \Big[\lambda\nabla^\mu \phi^I-\frac{1}{4}\left(\mathcal{K}(\nabla^\mu \phi^I) F^2\right)\Big]=0.\label{KG}
\end{gather}

Note that in holographic axion models, we have the following isotropic and homogeneous background metric,
\begin{gather} 
ds^2=\frac{1}{u^2}\left[-f(u)dt^2+\frac{1}{f(u)}du^2+dx^2+dy^2\right],\label{Background}\\
A=A_t(u)dt, \qquad \phi^I=\alpha \delta^I_i x^i,
\end{gather}
where $\alpha$ is a constant characterizing the breaking of translations, and $x^i$ denotes the boundary spatial coordinates. AdS boundary is located at $u=0$ and horizon is determined by $f(u=u_h)=0$ in this convention. For our model, we can solve the black hole solution analytically
\begin{gather}
f(u)=1-\frac{1}{2} \alpha^2 \lambda  u^2+\frac{\rho ^2 u^3 ArcTanh\left(\alpha \sqrt{-\mathcal{K}} u\right)}{4 \alpha \sqrt{-\mathcal{K}}}-\frac{\rho ^2 u^3 ArcTanh\left(\alpha \sqrt{-\mathcal{K}} u_h\right)}{4 \alpha \sqrt{-\mathcal{K}}}-\frac{u^3}{u_h^3}+\frac{\alpha^2 \lambda  u^3}{2 u_h},\\
A_t(u)=\mu -\frac{\rho  ArcTanh \left(\alpha \sqrt{-\mathcal{K}} u\right)}{\alpha \sqrt{-\mathcal{K}}}.
\end{gather}
Here, $\mu$ and $\rho$ are two integration constants which are  interpreted as chemical potential and charge density in the dual field theory. The regularity for gauge field $A_\mu$ on the horizon requires
\begin{equation}
\mu =\frac{\rho  ArcTanh \left(\alpha \sqrt{-\mathcal{K}} u_h\right)}{\alpha \sqrt{-\mathcal{K}}}.
\end{equation}
According to the holographic dictionary, entropy density, energy density, momentum susceptibility and temperature are respectively given by
\begin{gather}
s=\frac{4\pi}{u_h^2},\qquad \epsilon=2\left(\frac{1}{u_h^3}-\frac{\lambda \alpha^2}{2 u_h}+\frac{\rho^2 ArcTanh \left(\alpha\sqrt{-\mathcal{K}} u_h\right)}{4 \alpha \sqrt{-\mathcal{K}}}\right),\\
\chi_{PP}=\epsilon+P=\frac{3}{2}\epsilon, \qquad T=\frac{3}{4\pi u_h}-\frac{\lambda\alpha^2 u_h}{8\pi}-\frac{\rho^2 u_h^3}{16\pi(1+\mathcal{K}\alpha^2 u_h^2)},
\end{gather}
where $\chi_{PP}=\epsilon+P=3/2\epsilon$ comes from conformal invariance.

Next, we will explain the two different manners of breaking translations by analyzing the asymptotic behavior of scalar field $\phi^I$ close to the AdS boundary $(u\rightarrow 0)$. The story is very similar as in \cite{Amoretti:2016bxs,Alberte:2017oqx,Amoretti:2017frz,Amoretti:2017axe,Li:2018vrz,Amoretti:2019cef}.
\begin{itemize}
    \item If we set $\lambda \neq 0$ and $\mathcal{K}=0$ in (\ref{KG}), the asymptotic expansion of $\phi^I$ near the UV boundary is
\begin{equation}
\phi^I= \phi_{(0)}^I (t,x^i)+\phi_{(3)}^I (t,x^i)u^3+\cdots.
\end{equation}
According to the holographic dictionary and standard quantization, the leading order $\phi_{(0)}^I (t,x^i)=\alpha x^I \neq 0$ plays the role of external source that breaks translations explicitly, relaxing the momentum on the field theory side. Further properties and applications in holography have been widely investigated in \cite{Andrade:2013gsa,Davison:2014lua,Kim:2014bza,Kim:2015dna,Davison:2015bea,Blake:2018leo,Zhou:2019xzc}.

\item If we instead set $\lambda = 0$ and $\mathcal{K}\neq 0$, the asymptotic expansion of scalar field will be changed by
\begin{equation}
\phi^I=\frac{ \phi_{(-1)}^I (t,x^i)}{u}+\phi_{(0)}^I (t,x^i)+\cdots. \label{phiSSB}
\end{equation}     
In this case, the background solution of $\phi^I$ means that the leading $\phi_{(-1)}^I (t,x^i)$-term associated to the source should be turned off while existing a non-zero expectation value $<\mathcal{O}^I>$ $\sim$ $\phi_{(0)}^I (t,x^i)=\alpha x^I$. This implies the  pattern of spontaneous breaking of translational symmetry.
\end{itemize}

Note that if the translations are broken spontaneously, there should exist gapless excitations in the low energy description which are called Goldstone modes. In solid state physics, these Goldstone modes are called phonons, the propagating sound modes in elastic media. The dynamics of Goldstone modes has already been studied in holography \cite{Alberte:2017oqx,Ammon:2019wci,Ammon:2019apj,Baggioli:2019abx,Amoretti:2019cef,Donos:2019hpp,Amoretti:2019kuf,Baggioli:2020edn} and field theory \cite{Leutwyler:1996er,Dubovsky:2005xd,Nicolis:2015sra,Delacretaz:2017zxd,Alberte:2018doe,Musso:2018wbv}.

\subsection{Shear viscoelasticity}\label{SecElas}
One of the most distinctions of a solid and a fluid is that the former is resistant to shear deformations while the latter one is not.\footnote{However, in high momentum regime, fluids exhibit similar behaviors with solids. For a comprehensive review, one refers to \cite{Baggioli:2019jcm}.} To show that the black hole solution under consideration is dual to a solid system, we shall compute the Green function of the stress tensor, which in the hydrodynamic regime can be expressed as
\begin{equation}
\mathcal{G}^R_{T_{xy}T_{xy}}(\omega,k=0)=G\,-i\,\omega\,\eta+\cdots,
\end{equation}   
where $G$ is the shear elastic modulus and $\eta$ is the shear viscosity. In the rest of this subsection, we focus on the purely SSB pattern by turning off the external source, i.e, setting $\lambda=0$ and remaining $\mathcal{K}\neq0$. One can calculate $G$ and $\eta$ via retarded Green functions of stress tensor operator $T_{xy}$ numerically and read them by \cite{Ammon:2019wci}
\begin{equation}
G=Re\left[\mathcal{G}_{T_{xy}T_{xy}}^R(\omega=k=0)\right],\qquad \eta=-\lim_{\omega \rightarrow 0}\frac{1}{\omega}Im\left[\mathcal{G}_{T_{xy}T_{xy}}^R(\omega,k=0)\right].\label{Geta}
\end{equation}

\begin{figure}[htbp]
  \vspace{0.3cm}
  \centering
  \includegraphics[width=0.45\textwidth]{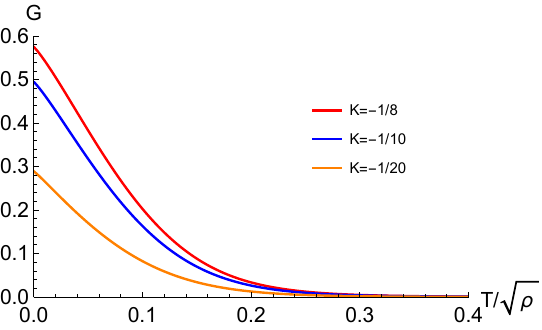} 
  \includegraphics[width=0.45\textwidth]{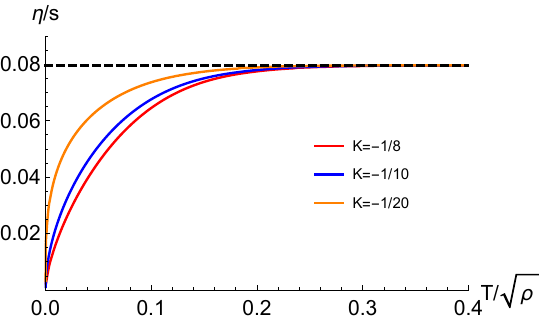}
  \caption{The dependence of the shear elastic modulus $G$ and the ratio of $\eta/s$ on $T/\sqrt{\rho}$ for different $\mathcal{K}$. The black dashed line is KSS bound $1/4\pi$ \cite{Kovtun:2004de}. We have fixed coupling $\lambda=0$.}\label{ModulusVis}
    \vspace{0.3cm}
\end{figure}

The numeric results have been shown in Fig.\ref{ModulusVis}. And we find that the ratio of shear viscosity $\eta$ to entropy density $s$ always breaks Kovtun-Son-Starinets (KSS) bound \cite{Kovtun:2004de} due to the breaking of translations. Moreover, there is a non-trivial shear modulus which implies the dual system is in a solid state. To have it positive, i.e avoiding dynamical instability, we should always require that $\mathcal{K}<0$ in our model.

Now, we analyze the Green function $\mathcal{G}_{T_{xy}T_{xy}}^R$ analytically in the high and low temperature limits. Let us focus on the former first. The details of the calculation are shown in the Appendix \ref{DeriDetail}. In the high temperature limit, the analytical results of shear modulus and viscosity are respectively given by
\begin{align}
&G=-\frac{9\mathcal{K} \alpha ^2 T}{128 \pi ^3}\left(\frac{\sqrt{\rho}}{T}\right)^4+\cdots, \label{AnalyG}\\
&\frac{\eta}{s}=\frac{1}{4 \pi }+\frac{81 \mathcal{K}\alpha^2}{16384 \pi ^7 T^2 }\left(\frac{\sqrt{\rho}}{T}\right)^4+\cdots.\label{AnalyEta}
\end{align}
Note that the positive shear modulus again requires the coupling constant $\mathcal{K}<0$. 

In the low temperature limit, we only find the analytical result for $\eta$ that is given by 
\begin{equation}
\frac{\eta}{s}\sim \left(\frac{T}{\sqrt{\rho}} \right)^{2\xi-1},\qquad \xi=\frac{1}{2} \sqrt{1-\frac{8 \sqrt{3} \mathcal{K}}{\sqrt{3 \mathcal{K}^2+1}}}.\label{xi2}
\end{equation}
This result is slightly different from the massive gravity without charge density \cite{Baggioli:2019rrs}, which shows that the ratio is proportional to $T^2$. The author of \cite{Hartnoll:2016tri} have claimed that $T^2$ was the fastest decay for the models whose IR geometry has an $AdS_2$ factor. In our case, the maximum of exponent is $T^2$ that is corresponding to $\mathcal{K}\rightarrow -\infty$, which satisfies this bound. However, this claim is not true anymore if one consider more complicated IR fixed points (see \cite{Ling:2016ien}). Note that for $\mathcal{K}=0$ the $\eta/s$ ratio is a constant which should be KSS bound. 

\section{Holographic transverse phonons}\label{lespectrum}

\subsection{Gapless phonons}\label{Gapless}

The dispersion relation of transverse phonons at low energy is
\begin{equation}
\omega=\pm c_T k-i D_T k^2+\cdots, \label{relation}
\end{equation}
where $c_T$ is the propagation velocity of phonons, which is related to the shear elastic modulus of material given by
\begin{equation}
c_T^2=\frac{G}{\chi_{PP}}.\label{SoundSpeed}
\end{equation}
$D_{T}$ is the attenuation constant that is related to the shear viscosity $\eta$ by $D_T=\eta/\chi_{PP}$ \cite{Amoretti:2019cef}. In order to confirm the presence of transverse phonons, we compute the quasi-normal modes (QNMs) numerically at zero and finite momenta in the infalling Eddington-Finkelstein coordinates (\ref{EF}) by using the pseudo-spectral method. It shows that the Goldstone modes induced by SSB of translations can be confirmed as transverse phonons in this holographic model (\ref{action}). One can see more numerical details in \cite{Jansen:2017oag}. \footnote{In the numerical setup throughout the paper, we fix that $u_h=1$ and $\alpha/\sqrt{\rho}=1$.}

\begin{figure}[htbp]
  \vspace{0.3cm}
  \centering
  \includegraphics[width=0.42\textwidth]{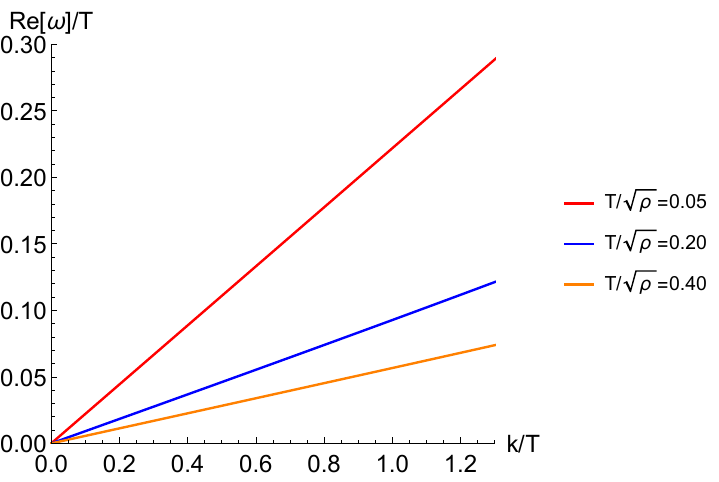}
  \includegraphics[width=0.47\textwidth]{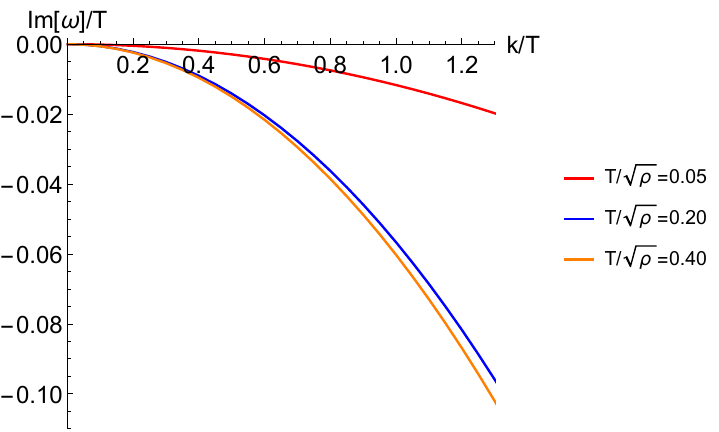}
  \caption{Real and imaginary parts of the lowest QNMs for different temperatures. We have fixed coupling $\mathcal{K}=-1/8$ and $\lambda=0$. We try to use the generalized dispersion relation $\omega=c_1 k^{a}+ic_2 k^{b}$ to fit the QNMs data. In our numerical cases, we get $a=1.00003\pm 0.00002$ and $b=2.00003\pm 0.00003$, which satisfies the dispersion relation (\ref{relation}).}\label{QNM1}
    \vspace{0.3cm}
\end{figure}

\begin{figure}[htbp]
  \vspace{0.3cm}
  \centering
  \includegraphics[width=0.42\textwidth]{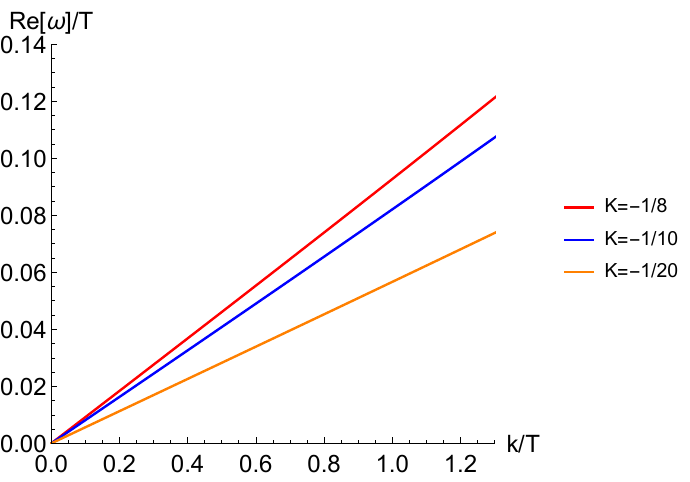}
  \includegraphics[width=0.47\textwidth]{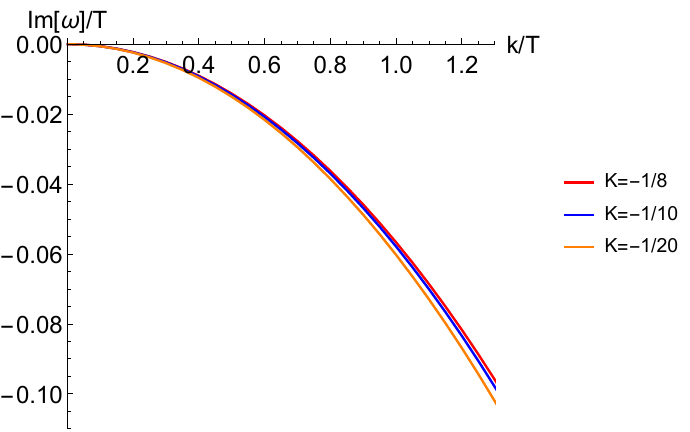}
  \caption{Real and imaginary parts of the lowest QNMs for different couplings $\mathcal{K}$. We have fixed $T/\sqrt{\rho}=0.2$ and $\lambda=0$. }\label{QNM2}
    \vspace{0.3cm}
\end{figure}

\begin{figure}[htbp]
  \vspace{0.3cm}
  \centering
  \includegraphics[width=0.48\textwidth]{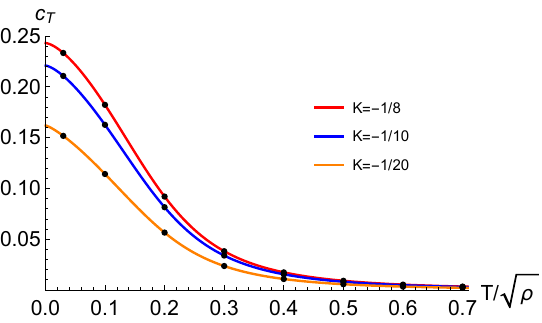}
  \caption{Comparison of the sound velocity extracted from QNMs (black dots) and the result from the elasticity formula (\ref{SoundSpeed}) (solid lines) for various values of $\mathcal{K}$. }\label{ShearModulus}
    \vspace{0.3cm}
\end{figure}

In Fig.\ref{QNM1}, we show the real and imaginary parts of the lowest order QNMs for different temperatures. These Goldstone modes are phonons we expected. The spectrums of QNMs for different couplings are shown in Fig.\ref{QNM2}. One can calculate the sound speed of the transverse phonons in terms of (\ref{SoundSpeed}) by holography and compare them to the numeric results extracted from QNMs. We find that they are in perfect agreement with each other, as is shown in Fig.\ref{ShearModulus}.

\subsection{Pinned phonons}\label{pinnedphonons}
In this section, we further introduce slow relaxation on momentum into the SSB pattern. Physically, this makes the momentum operator no longer exactly conserved and the would be massless Goldstones get slightly gapped (which is now called pseudo-Goldstones or pinned phonons). In our model it can be easily realized by including the canonical kinetic term of the axions as follows,
\begin{align}
\mathcal{L}_{\phi}&=-\lambda X-\frac{1}{4}\mathcal{K} X F ^2.\label{ScalarAction}
\end{align}
In order not to change the SSB pattern significantly, we assume that the value of $\lambda$ is small comparing to $\mathcal{K}$.

In presence of slow momentum relaxation, the relaxation rate $\Gamma$ is controlled in the manner of \cite{Andrade:2013gsa,Davison:2013jba}
\begin{equation}
\Gamma \propto   \lambda\,\alpha^2.
\end{equation}
Therefore we can define a ESB scale which depicts the strength of explicit breaking of symmetry
\begin{equation}
\langle\textsl{ESB}\rangle\equiv \sqrt{\lambda}\,\alpha. \label{DeESB}
\end{equation}
Heuristically, we define the SSB scale as follows
\begin{equation}
\langle\textsl{SSB}\rangle\equiv \sqrt{\frac{-\mathcal{K}}{4}}\,\alpha.\label{DeSSB}
\end{equation}
Then, the competition of SSB and ESB should be described by a dimensionless parameter that is defined by
\begin{equation}
\beta\equiv \frac{\langle \textsl{SSB} \rangle}{\langle \textsl{ESB} \rangle}=\sqrt{\frac{-\mathcal{K}}{4\,\lambda}}.
\end{equation}
That is to say that at $\beta=0$ the breaking of translations is totally explicit while at $\beta=\infty$ purely spontaneous. Then, the pinned phonons are expected to appear when $\beta\gg 1$ (pseudo-spontaneous). In this case, the hydrodynamic formulation predicts the following universal result for zero momentum \cite{Delacretaz:2017zxd}
\begin{equation}
\left(\Omega-i\omega) (\Gamma-i\omega\right)+\omega_0^2=0,\label{HyfroOmega}
\end{equation}
where $\Gamma$ is the momentum relaxation rate (charactering how fast the total momentum is dissipated), $\Omega$ is the phase relaxation rate (measuring the lifetime of the pinned phonons) and $\omega_0$ is the so-called pinning frequency (the mass of the pinned phonons). Solving the equation (\ref{HyfroOmega}) we obtain a pair of modes
\begin{equation}
\omega_\pm=-\frac{i}{2}\left(\Omega+\Gamma\right)\pm \frac{1}{2}\sqrt{4\omega_0^2-\left(\Gamma-\Omega\right)^2}.\label{HyfroOmega1}
\end{equation}
Note that this formula is valid only when the two modes enter in the hydrodynamic limit $|\omega_\pm| /T\ll 1$.

\begin{figure}[htbp]
  \vspace{0.3cm}
  \centering
  \includegraphics[width=0.51\textwidth]{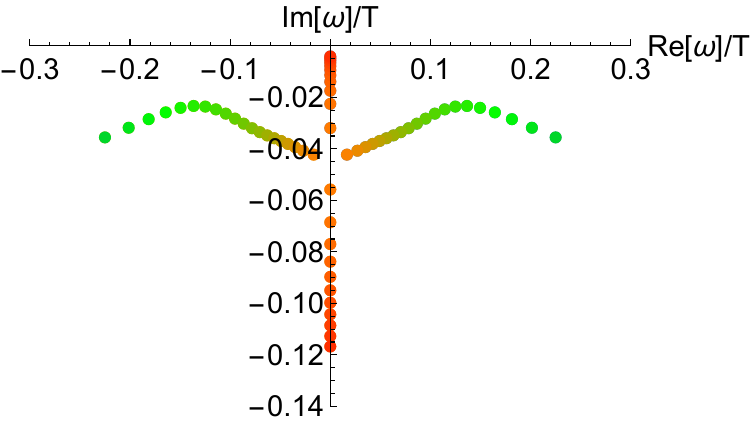}\hspace{5mm}
  \includegraphics[width=0.42\textwidth]{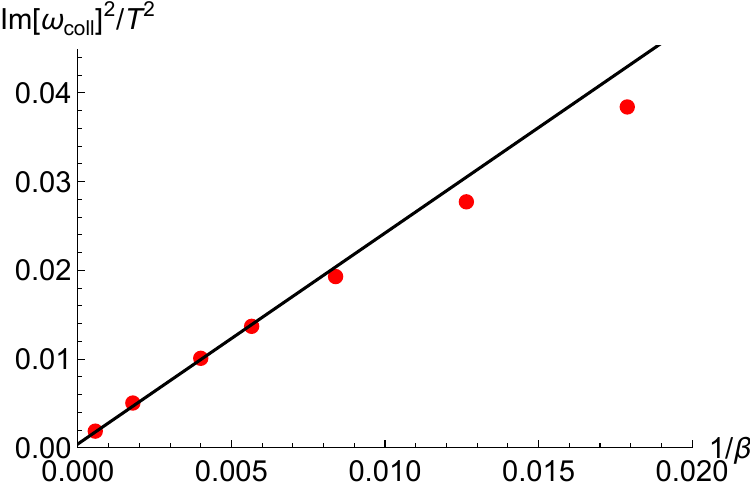}
  \caption{{\bf Left}: The lowest two QNMs at zero momentum for $\mathcal{K}=-1/8$. We have fixed $\lambda=1\times 10^{-8}$ and $T/\sqrt{\rho}$ $\in$ [0.1, 0.4](green-red). {\bf Right}: Collision frequency $Im \,\omega_{coll}$ as the function of $\beta=\langle SSB \rangle/\langle ESB \rangle$. We have fixed $\mathcal{K}=-1/8$, $\lambda$ $\in$ $[1\times 10^{-8}, 1\times 10^{-5}]$ and $\alpha/\sqrt{\rho}=1$.  The black line indicates linear scaling.}\label{collision}
    \vspace{0.3cm}
\end{figure}

Next we show the QNMs and see how these two modes move. For this purpose we fix $\mathcal{K}$ and $\lambda$ and tune the temperature $T/\sqrt{\rho}$. The results are shown in the left panel of Fig.\ref{collision}. We find that at high temperatures (i.e, small values of $\langle ESB\rangle/T$) the two modes lie on the imaginary axes. In this case, we have $4\omega_0^2 <\left(\Gamma-\Omega\right)^2$. When decreasing temperature $4\omega_0^2$ grows faster than $(\Gamma-\Omega)^2$, the two modes get closer, collide and finally become off-axis.

In condensed matter systems, the presence of the phase relaxation could be attributed to either the proliferation of dislocations \cite{Delacretaz:2017zxd} or the disorder effects \cite{phase1,phase2}. And in holographic models, it is associated with the collision mechanism of the two lowest-lying quasi-normal modes \cite{Ammon:2019wci}. It is worth studying the relationship between collision frequency $\omega_{coll}$ and characteristic parameter $\beta$. We can fix $\mathcal{K}$, choose one specific $\lambda$ and show the QNMs spectrum. Then identify the collision frequency at the collision point. The final results are shown in the right panel of Fig.\ref{collision}. For $\beta\gg 1$ the collision happens towards the origin so that very close to the origin, which can be described effectively by hydrodynamics. When $\beta=\infty$ (purely SSB case), the two modes are all located at the origin. As shown in the right panel of Fig.\ref{collision}, the collision frequency $\omega_{coll}$ shows proportional to the characteristic parameter $1/\beta$ approximately, that is
\begin{equation}
\frac{Im\,\omega_{coll}}{T}\sim \sqrt{\frac{\langle ESB \rangle}{\langle SSB \rangle}}.
\end{equation}

\begin{figure}[htbp]
  \vspace{0.3cm}
  \centering
  \includegraphics[width=0.47\textwidth]{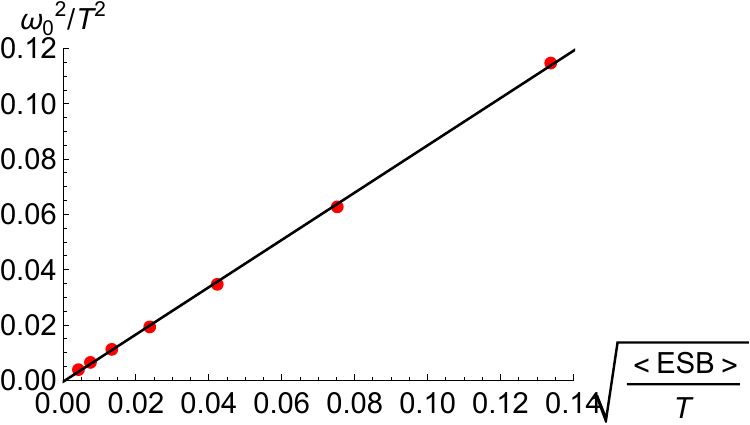}\hspace{5mm}
  \includegraphics[width=0.49\textwidth]{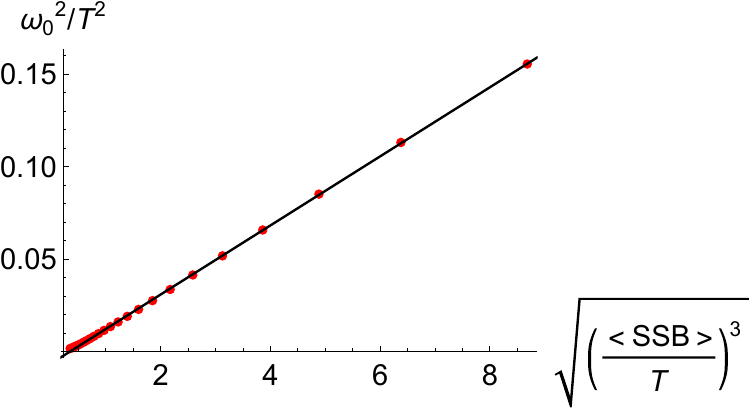}
  \caption{The dependence of the pinning frequency $\omega_0$ on the ESB and SSB scales. {\bf Left}: We have fixed $\lambda$ $\in$ $[1\times 10^{-11}, 1\times 10^{-5}]$,  $T/\sqrt{\rho}=0.15$, $\alpha/\sqrt{\rho}=1$ and $\mathcal{K}=-1/8$. {\bf Right}: We have fixed $T/\sqrt{\rho}\in[0.06, 0.30]$, $\alpha/\sqrt{\rho}=1$, $\lambda=1\times 10^{-8}$ and $\mathcal{K}=-1/8$. The black lines indicate linear scalings.}\label{GMOR}
    \vspace{0.3cm}
\end{figure}

Next we check Gell-Mann-Oakes-Renner (GMOR) relation which was originally  found in QCD \cite{GellMann:1968rz} and later was also checked in holographic studies \cite{Kruczenski:2003uq,Erlich:2005qh,Filev:2009xp,Argurio:2015wgr,Amoretti:2016bxs,Ammon:2019wci,Baggioli:2019abx}. In the pseudo-spontaneous limit, the momentum relaxation rate can be neglected in (\ref{HyfroOmega1}). By reading QNMs data, phase relaxation rate $\Omega$ and pinning frequency $\omega_0$ can be identified. We show the dependence of the pinning frequency $\omega_0$ on the ESB and SSB scales in Fig.\ref{GMOR}. We find that the pinning frequency $\omega^2_0$ is proportional to the square root of the explicit breaking scale (\ref{DeESB}) and to the half to the third power of the spontaneous breaking scale (\ref{DeSSB}). That said, in the pseudo-spontaneous limit, the results indicate the following scaling behavior
\begin{equation}
\omega^2_0\sim \sqrt{\langle ESB \rangle\langle SSB \rangle^3},
\end{equation}
which is different from the original GMOR relation for Pion. Similarly in \cite{Andrade:2017cnc,Andrade:2018gqk,Andrade:2020hpu}, they also obtained the results which is different from GMOR relation in holography.\footnote{Note that the derivation of the  original GMOR relation was based on a specific QCD model. To our best knowledge, there is by now no rigorous proof that this relation should be valid for all cases in the field theory side. Holographic models suggest a more general form which is ${\omega_0^2}\sim \langle ESB \rangle^a\langle SSB \rangle^b$ with two non-negative exponents $a$ and $b$. For example in holographic model with helical structure, the exponents are $a=2$ and $b=2$ \cite{Andrade:2017cnc}.} However in a simple holographic axion model, the authors proved this conclusion \cite{Ammon:2019wci}.

\section{Electric conductivity}\label{ctransport}
When the charge sector strongly couples with phonons, the combined system will reach local equilibrium extremally fast. Then, total momentum of the charge-phonon system should be conserved, resulting an infinite DC conductivity.\footnote{Here, we have assumed that the system has no particle-hole symmetry.} Such a divergency of the conductivity can again be removed by introducing an ESB of translations which relaxes the momentum. However, unlike the picture without phonons, the AC conductivity now displays a finite peak at finite frequency. 
In this section, we turn to check how (pseudo-)phonons in our holographic model affect the charge transport in the purely SSB case and in the presence of weakly and explicitly broken translations, respectively.

\subsection{Purely SSB pattern}
In the presence of purely SSB case, the AC conductivity at low frequency obeys the following hydrodynamic formula \cite{Hartnoll:2016apf}
\begin{equation}
\sigma(\omega)=\sigma_{Q}+\frac{\rho^2}{\chi_{PP}}\left(\delta (\omega)+\frac{i}{\omega}\right),\label{incoherent1}
\end{equation}
where $\sigma_{Q}$ is the so-called incoherent conductivity and $\rho^2/\chi_{PP}$ is Drude weight. Note that the DC conductivity is infinite  in the SSB case because of the absence momentum relaxation mechanism. In this holographic model, the incoherent conductivity can be computed directly via the membrane paradigm (see \cite{Davison:2015taa} for more details)
\begin{equation}
\sigma_{Q}=\left(1+\mathcal{K}\frac{4\pi\alpha^2}{s} \right)\left(\frac{sT}{sT+\mu\rho}\right)^2.\label{incoherent2}
\end{equation}
Here, the first law of thermodynamics $\epsilon+p=sT+\mu\rho$ has been applied.

We follow the procedure in \cite{Andrade:2013gsa} to calculate AC conductivity at zero momentum by holographic method. Introduce the time-dependent perturbations around the background (\ref{Background})
\begin{equation}
\delta g_{tx}=\frac{1}{u^2}h_{tx}(u)e^{-i\omega t}, \quad \delta A_{x}= a_x(u)e^{-i\omega t},\quad \delta \phi_{x}=\phi_{x}(u)e^{-i\omega t},
\end{equation}
and choose radial gauge $\delta g_{ux}=0$.
The linearized equations of motion are given by
\begin{align}
&0=2 f (\alpha ^2 u^2 \mathcal{K}+1)^2 (\rho  u^3 a_x'-u h_{tx}''+2 h_{tx}')+\alpha ^2 u h_{tx} (2 \lambda  (\alpha ^2 u^2 \mathcal{K}+1)^2-\rho ^2 u^4 \mathcal{K})
\nonumber\\&
+i \alpha  u \phi_{x}  \omega  (2 \lambda  (\alpha ^2 u^2 \mathcal{K}+1)^2-\rho ^2 u^4 \mathcal{K}),\label{ACEQ1}\\
&0=2 i \rho  \omega  a_x (\alpha ^2 u^3 \mathcal{K}+u)^2+\alpha  f \phi_{x} ' (2 \lambda  (\alpha ^2 u^2 \mathcal{K}+1)^2-\rho ^2 u^4 \mathcal{K})-2 i \omega  h_{tx}' (\alpha ^2 u^2 \mathcal{K}+1)^2,\\
&0=\omega ^2 a_x (-(\alpha ^2 u^2 \mathcal{K}+1))-f (a_x' (\alpha ^2 u \mathcal{K}(u f'+2 f)+f')+f a_x'' (\alpha ^2 u^2 \mathcal{K}+1)-\rho  h_{tx}'),\\
&0=f (\phi_{x} ' (2 f (2 \lambda  (\alpha ^2 u^2 \mathcal{K}+1)^3+\rho ^2 u^4 \mathcal{K} (1-\alpha ^2 u^2 \mathcal{K}))-u f' (\alpha ^2 u^2 \mathcal{K}+1) (2 \lambda  (\alpha ^2 u^2 \mathcal{K}+1)^2
\nonumber\\&
-\rho ^2 u^4 \mathcal{K}))-f u \phi_{x} '' (\alpha ^2 u^2 \mathcal{K}+1)(2 \lambda  (\alpha ^2 u^2 \mathcal{K}+1)^2-\rho ^2 u^4 \mathcal{K}))-u \phi_{x}  \omega ^2 (\alpha ^2 u^2 \mathcal{K}+1) (2 \lambda  (\alpha ^2 u^2 \mathcal{K}+1)^2
\nonumber\\&
-\rho ^2 u^4 \mathcal{K})+i \alpha  u \omega  h_{tx} (\alpha ^2 u^2 \mathcal{K}+1) (2 \lambda  (\alpha ^2 u^2 \mathcal{K}+1)^2-\rho ^2 u^4 \mathcal{K}).\label{ACEQ2}
\end{align}
It is easy to check there are three independent equations above. One can reduce them to two equations (see \cite{Wang:2019jyw} for analogous simplification method). The AC conductivity can be computed by solving (\ref{ACEQ1}-\ref{ACEQ2}) numerically via Kubo formula \cite{Ammon:2019wci}
\begin{equation}
\sigma(\omega)=\frac{1}{i\omega} \mathcal{G}_{JJ}^R(\omega,k=0).\label{AC}
\end{equation}

\begin{figure}[htbp]
  \vspace{0.3cm}
  \centering
  \includegraphics[width=0.47\textwidth]{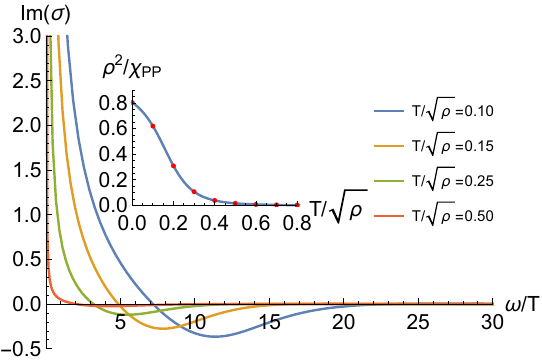}
  \includegraphics[width=0.48\textwidth]{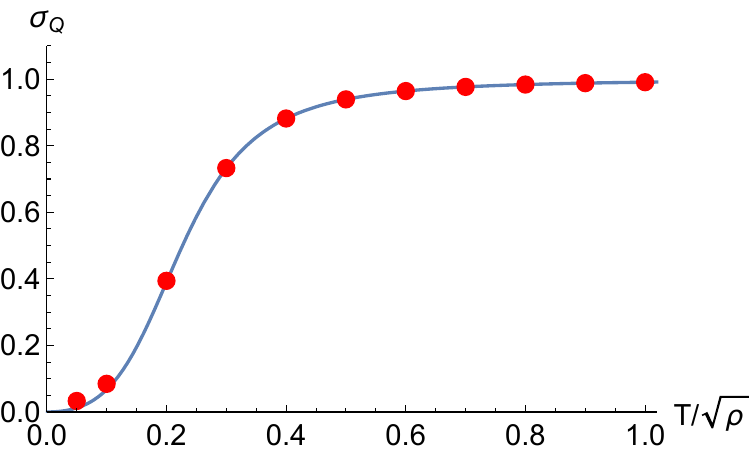}
  \caption{{\bf Left}: The imaginary parts of AC conductivity for different temperatures. The inset plot shows the Drude weight $\rho^2/\chi_{PP}$. Blue line is the analytical result obtained from (\ref{incoherent1}) and red dots are the numerical results from (\ref{AC}). {\bf Right}: The incoherent conductivity $\sigma_Q$ as the function of temperature. Blue line is the analytical result obtained from (\ref{incoherent2}) and red dots are the numerical results from (\ref{AC}) . We have fixed $\mathcal{K}=-1/8$ and $\lambda=0$.}\label{ImIncoherent}
    \vspace{0.3cm}
\end{figure}

We show that the numerical results of AC and incoherent conductivities in Fig.\ref{ImIncoherent}. These results imply that the background solution of scalar field $\phi^I=\alpha x^I$ should be interpreted as the SSB nature. Moreover, the incoherent conductivity is also controlled by $\alpha$ and coupling $\mathcal{K}$, which can be viewed as the effect from phonons. Note that this can never happen if we neglect the direct coupling the gauge field and the axions in the bulk theory. It is obvious to see in Fig.\ref{ImIncoherent} that our numerical results from (\ref{AC}) are in agreement with analytical results from (\ref{incoherent1}-\ref{incoherent2}) .

\subsection{Pinned structure}
\begin{figure}[htbp]
  \vspace{0.3cm}
  \centering
  \includegraphics[width=0.47\textwidth]{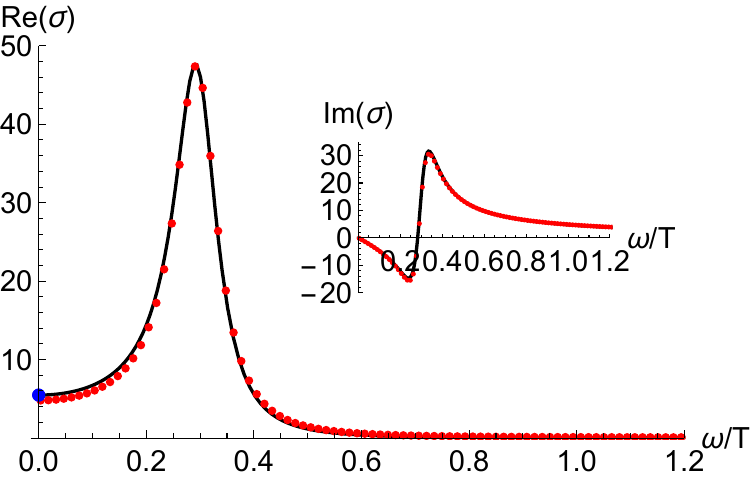}
  \includegraphics[width=0.47\textwidth]{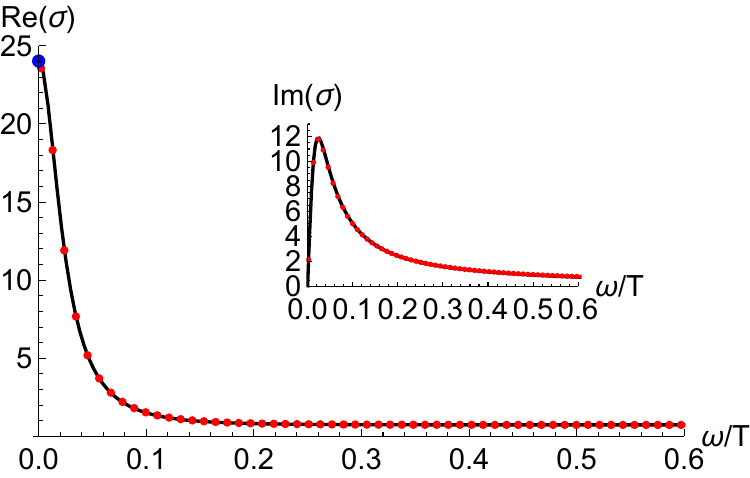}
  \caption{Comparisons between numerical AC conductivities (black solid lines) and hydrodynamic predictions (red dots) for different temperatures. The blue dots are the DC conductivity results from (\ref{DCformula}). {\bf Left}: $T/\sqrt{\rho}=0.1$ and $\omega_0^2\approx 0.0017$. Here, one should distinguish $\omega_0$-the mass of pinned phonons and $\omega_{\text{peak}}$, the peak of the conductivity in IR region. {\bf Right}: $T/\sqrt{\rho}=0.3$. We have fixed $\lambda=1\times 10^{-7}$ and $\mathcal{K}=-1/8$.}\label{QNMsAC}
    \vspace{0.3cm}
\end{figure}

In section \ref{pinnedphonons}, we study the pinned phonons by turning on small external source. In this case, the DC conductivity can be computed anatically by the membrane paradigm \cite{Donos:2014cya,Baggioli:2016pia}
\begin{equation}
\sigma_{DC}=1+\mathcal{K} \alpha^2  u_h^2+\frac{\rho^2 u_h^2}{\alpha^2\left(\lambda-\frac{\mathcal{K}\rho^2 u_h^4}{2\left(1+\mathcal{K} \alpha^2  u_h^2\right)^2}\right)}.\label{DCformula}
\end{equation}
The structure of the pinned phonons are also captured by the AC conductivity in the low frequency. AC conductivity (\ref{incoherent1}) from hydrodynamics should be modified in the presence of both ESB and SSB, which is given by \cite{Amoretti:2019kuf,Amoretti:2018tzw}
\begin{equation}
\sigma(\omega)=\sigma_Q+\frac{\gamma_1^2\chi_{PP}^2\omega_0^2(\Gamma-i\omega)+2\gamma_1 \rho \chi_{PP}\omega_0^2+\rho^2(\Omega-i\omega)}{\chi_{PP}^2(\left(\Omega-i\omega) (\Gamma-i\omega\right)+\omega_0^2)}.\label{pinnedAC}
\end{equation}
where $\gamma_1$ is the coefficient which is extracted from the correlator between the electric current and Goldstone operator. When the momentum relaxation is slow, relaxation rate $\Gamma$ can be ignored. Based on our calculation, $\omega_0^2$ is the very small quantity. Thus we can omit the contribution of the $\gamma_1$ term. We identify $\Omega$ and $\omega_0$ by extracting QNMs data, which will fix all the parameters (\ref{pinnedAC}), to fit the AC conductivity by holography. The results are shown in Fig.\ref{QNMsAC}. We observe that the numerical AC conductivities are fitted by the hydrodynamic predictions well. As the authors point out \cite{Baggioli:2016oqk,Gouteraux:2016wxj,An:2020tkn}, the pinning effect introduces the metal-insulator transition. From the perspective of the shear viscosity, there is no specific behaviour at both sides of the transition point (see Appendix \ref{DCVis} for more details).

\section{Conclusions}\label{conclusion}
In this work, we consider a holographic model that can break translational invariance in the dual field theory. We confirm the explicit breaking by turning on the linear axion term $\lambda X$ and spontaneous breaking by a gauge-axion coupling $\mathcal{K} XF^2$ via the UV analysis.

When turning off the external source, we compute shear modulus and viscosity by holography. The non-trivial results imply that the dual boundary system is an elastic solid. By calculating quasi-normal modes and making a comparison with the hydrodynamics predictions, we verify the existence of the gapless transverse phonons in the low energy spectrum. Then we turn on the slow momentum relaxation to see pinned phonons and check GMOR relation. It is found that the pinning frequency is proportional to the square root of the explicit breaking scale and to the half to the third power of the spontaneous breaking scale. This implies that it is different from the ``standard" GMOR relation in our model.

We then study the charge transport. Firstly in the purely SSB pattern, the AC conductivity is in perfect agreement with the hydrodynamics expectations. This allows us to claim that the background solution of scalar field should be interpreted as the SSB contribution. Meanwhile in the presence of weakly explicit breaking, we show that the numerical AC conductivity can also be fitted by the hydrodynamic predictions well. 

In the small frequency and momentum regime, our model shares the same hydrodynamic formulas with other holographic axion models \cite{Baggioli00:2021xuv} or the homogeneous models with more complicated setups \cite{Amoretti:2018tzw,Andrade:2017cnc} (even though the coefficients and parameters in the hydrodynamic formulation depend on the setup). This again verifies the statement that hydrodynamics provides a universal description for strongly coupled systems.

In this paper we only focus on the transverse channel of the system. The spectra of the longitudinal channel are also worth analyzing. This channel contains a pair of sound modes and one crystal diffusion mode in the low momentum limit, which is more complicated \cite{Ammon:2019apj,Baggioli:2019aqf,Baggioli:2019abx}. We can study how charge density affects speed of longitudinal sounds, crystal diffusion and  check the universality of its lower \cite{Baggioli:2020ljz,Wu:2021mkk} and upper \cite{Baggioli:2020ljz,Arean:2020eus,Wu:2021mkk} bounds proposed previously. In addition, one can also investigate the magnetic effect with the current model and compare the result with that from the magneto-hydrodynamics \cite{Amoretti:2019buu,Delacretaz:2019wzh}  and other holographic models with magnetic effects \cite{Baggioli:2020edn,Amoretti:2020mkp,Amoretti:2021fch,Donos:2021ueh}. We will leave these for future work.

\acknowledgments
We would like to thank Andrea Amoretti, Matteo Baggioli, Guo-Yang Fu, Li Li, Yi Ling, Jian-Pin Wu and Meng-He Wu for innumerous stimulating discussions and comments on the topics presented in this paper. We are particularly grateful to Matteo Baggioli and Li Li for their valuable comments about the manuscript. XJW is supported by NSFC No.11775036 and the Postgraduate Research \& Practice Innovation Program of Jiangsu Province (KYCX21\_3192). WJL is supported by NSFC No.11905024 and No.DUT19LK20.

\appendix

\section{Derivation details for analytical Green's functions}\label{DeriDetail}

In this appendix, we show the details for the analytical calculations of Green's functions $\mathcal{G}_{T_{xy}T_{xy}}^R$ at high and zero temperature limits.

\subsection{High temperature limit}

Let us focus on the high temperature limit first, i.e. $\sqrt{\rho}/T\ll 1$ that is equivalent to small charge density limit. We use Poincar\'e coordinates (\ref{Background}) this time and turn on shear perturbation $\delta g_{xy}=e^{-i\omega t}h(u)/u^2$. The  perturbation equation is given by
\begin{equation}
h''+\left(\frac{f'}{f}-\frac{2}{u}\right)h'+\left(\frac{\omega ^2}{f^2}+\frac{\alpha ^2 \hat{\rho} ^2 T^4 u^4 \mathcal{K}}{2 f \left(\alpha ^2 u^2 \mathcal{K}+1\right)^2}\right)h=0, \label{ShearEq}
\end{equation}
where we have introduced a dimensionless parameter which is defined by
\begin{equation}
\hat{\rho}=\left(\frac{\sqrt{\rho}}{T}\right)^2.
\end{equation}

First, we show the analytical procedure for the correlator (\ref{Geta}). To solve the equation of motion (\ref{ShearEq}), one should implement the infalling boundary condition at the horizon and require field $h(u=0)=h_0$ at the AdS boundary in zero frequency limit. The on-shell action can be obtained \cite{Alberte:2016xja}
\begin{equation}
S_{bdy}=\int d^2 x\int \frac{d\omega}{2\pi} h_0 \mathcal{F}(\omega,u)h_0\bigg|_{u=\epsilon},\label{OnShell}
\end{equation}
where $\epsilon$ is the UV cut-off and $h_0$ is the value of the bulk field $h$ at the boundary, which is equivalent to the source $h_0\equiv h(u=0).$ As illustrated in \cite{Son:2002sd}, the retarded Green's function $\mathcal{G}^R_{T_{ij}{T_{ij}}}$ can be computed by taking two functional derivative of the above on-shell action with respect to the source $h_0$
\begin{equation}
\mathcal{G}^R_{T_{ij}{T_{ij}}}=-\lim_{\epsilon\rightarrow 0}2 \mathcal{F}(\omega,u)\bigg|_{u=\epsilon}.\label{derivative}
\end{equation}

The equation (\ref{ShearEq}) can be solved by imposing the follow ansatz
\begin{equation}
h(u)=h_0 \,e^{-\frac{i \omega}{4\pi T} log f}\left(\Phi_0(u)+\frac{i\omega}{4\pi T}\Phi_1(u)+\cdots\right),\label{hAnsatz}
\end{equation}
where exponential prefactor represents infalling behaviour and ellipsis denotes higher frequency contributions which are neglected. The functions $\Phi_0$ and $\Phi_1$ are required to be regular. We set 
\begin{equation}
\Phi_0(0)=1,\qquad \Phi_1(0)=0,\label{PhiPsi}
\end{equation}
near the boundary.

One can use perturbative method to solve the equation (\ref{ShearEq}) with the ansatz (\ref{hAnsatz}). The equations for $\Phi_0$ and $\Phi_1$ can be obtained order by order in the frequency
\begin{align}
&0=\left(\frac{f}{u^2}\Phi_0'(u)\right)'+\frac{\alpha ^2 \hat{\rho} ^2 T^4 u^2  \mathcal{K}}{2 \left(\alpha ^2 u^2 \mathcal{K}+1\right)^2}\Phi _0(u),
\nonumber\\
&0=\left(\frac{f}{u^2}\Phi_1'(u)\right)'+\frac{\alpha ^2 \hat{\rho} ^2 T^4 u^2  \mathcal{K}}{2 \left(\alpha ^2 u^2 \mathcal{K}+1\right)^2}\Phi _1(u)-\frac{2 f'}{u^2}\Phi _0'(u)+\left(\frac{2 f'}{u^3}-\frac{f''}{u^2}\right)\Phi _0(u).\label{OrderBy}
\end{align}

Since we consider small charge density limit, the emblackening factor $f$ can be simplified to Schwarzchild-AdS geometry $f(u)=1-u^3/u_h^3$ and the Hawking temperature is $T=3/(4\pi u_h)$. Meanwhile, we adopt the following series expansions for the ansatz (\ref{hAnsatz}) in $\hat{\rho}^2$
\begin{equation}
\Phi_0=\sum_{n=0}^\infty \hat{\rho}^{2n}\phi_n, \qquad \Phi_1=\sum_{n=0}^\infty \hat{\rho}^{2n}\psi_n,
\end{equation}
where the requirements (\ref{PhiPsi}) for $\Phi_0$ and $\Phi_1$ make $\phi_0=1$ and $\psi_0=0$. Substitute the above expansions into (\ref{OrderBy}), and one can obtain
\begin{align}
&\left(\frac{f}{u^2}\phi_1'\right)'=-\frac{\alpha ^2 u^2 \mathcal{K}T^4}{2 \left(\alpha ^2 u^2 \mathcal{K}+1\right)^2}\approx -\frac{1}{2} \alpha ^2  u^2 \mathcal{K} T^4,\label{Order1}
\\
&\left(\frac{f}{u^2}\psi_1'\right)'=2\frac{f'}{u^2}\phi_1'.\label{Order2}
\end{align}
By dimensional analysis, the dimension of $\alpha$ is the same as $\sqrt{\rho}$. We can expand $\alpha$ to get the leading order 
contribution, i.e. approximate term in (\ref{Order1}).

The on-shell action is given by \cite{Alberte:2016xja} \footnote{In this paper \cite{Alberte:2016xja}, the author re-defined the radial coordinate $z=u/u_h$. We recover the coordinate $u$ here.} 
\begin{equation}
S_{bdy}=-\frac{1}{2}\int d^3 x \left(\frac{f}{u^2}h(u)h'(u)\right)\bigg|_{u=\epsilon}^{u=uh},
\end{equation}
where $\epsilon$ is the UV cut-off. Expand the above formula at leading order in the charge density and frequency
\begin{equation}
S_{bdy}=-\int d^3 x\,\frac{h_0^2}{2}\left[\hat{\rho}^2\left(\frac{f}{\epsilon ^2}\phi _1'\right)+\frac{i\omega}{4\pi T}\left(\frac{3}{u_h^3}+\hat{\rho}^2\left(\frac{f}{\epsilon^2}\psi_1'\right)-2\hat{\rho}^2 log \,f\left(\frac{f}{\epsilon^2}\phi _1'\right)\right)+\cdots\right].\label{FinallOnShell}
\end{equation}
The shear elastic modulus and viscosity can be calculated according to (\ref{OnShell}-\ref{derivative}) and (\ref{FinallOnShell})
\begin{align}
&G=Re \mathcal{G}_{T_{xy}T_{xy}}^R= \lim_{\epsilon\rightarrow 0}\hat{\rho}^2\left(\frac{f}{\epsilon^2}\phi_1'\right),
\\
&\eta=\lim_{\omega\rightarrow 0}\left[-\frac{1}{\omega}Im \mathcal{G}_{T_{xy}T_{xy}}^R(\omega)\right]= \lim_{\epsilon\rightarrow 0}\frac{1}{4\pi T}\left(\frac{3}{u_h^3}+\hat{\rho}^2\left(\frac{f}{\epsilon^2}\psi_1'\right)\right).
\end{align}

Firstly, we compute the shear elastic modulus concretely by using (\ref{Order1})
\begin{align}
&G=\hat{\rho}^2\lim_{\epsilon \rightarrow 0}\int^{u_h}_{\epsilon}\left(-\frac{1}{2} \alpha ^2  u^2 \mathcal{K} T^4\right)\nonumber\\
&\,\,\,\,\,=-\frac{1}{6} \alpha ^2 \hat{\rho} ^2 T^4 \mathcal{K} u_h^3.
\end{align}
We recover the parameter $\rho$ and replace $u_h$ with $T$. Then one can re-write the above relation
\begin{equation}
G=-\frac{9\mathcal{K} \alpha ^2 T}{128 \pi ^3}\left(\frac{\sqrt{\rho}}{T}\right)^4+\cdots.\label{AnalyG2}
\end{equation}

Secondly, we show the shear viscosity. In small charge density limit, the equation (\ref{Order1}) for $\phi_1$ is solved by
\begin{equation}
\frac{f}{u^2}\phi_1'=\frac{1}{6} \alpha ^2 T^4 \mathcal{K} \left(u^3-u_h^3\right),\label{Order3}
\end{equation}
For (\ref{Order2}) and (\ref{Order3}),
\begin{equation}
\frac{f}{u^2}\psi_1'=\int_u^{u_h} 2\frac{f'}{x^2}\phi_1' dx=\int_u^{u_h} 2\frac{f'}{f}\left(\frac{f}{x^2}\phi_1'\right) dx=\frac{1}{3} \alpha ^2 T^4 \mathcal{K} \left(u_h^3-u^3\right).
\end{equation}
Thus the ratio of shear viscosity $\eta$ to entropy density $s$ is
\begin{align}
\frac{\eta}{s} &=\lim_{\epsilon\rightarrow 0}\frac{1}{4\pi T s}\left(\frac{3}{u_h^3}+\hat{\rho}^2\left(\frac{f}{\epsilon^2}\psi_1'\right)\right)\\
               &=\frac{1}{4 \pi }+\frac{81 \mathcal{K}\alpha ^2 \rho ^2 }{16384 \pi ^7 T^6},
\end{align}
where we have used simple emblackening factor $f(u)=1-u^3/u_h^3$, temperature $T=3/(4\pi u_h)$ and entropy density $s=4\pi/u_h^2$. Thus the above formula can be re-written
\begin{equation}
\frac{\eta}{s}=\frac{1}{4 \pi }+\frac{81 \mathcal{K}\alpha^2}{16384 \pi ^7 T^2 }\left(\frac{\sqrt{\rho}}{T}\right)^4+\cdots.\label{AnalyEta2}
\end{equation}

\begin{figure}[htbp]
  \vspace{0.3cm}
  \centering
  \includegraphics[width=0.45\textwidth]{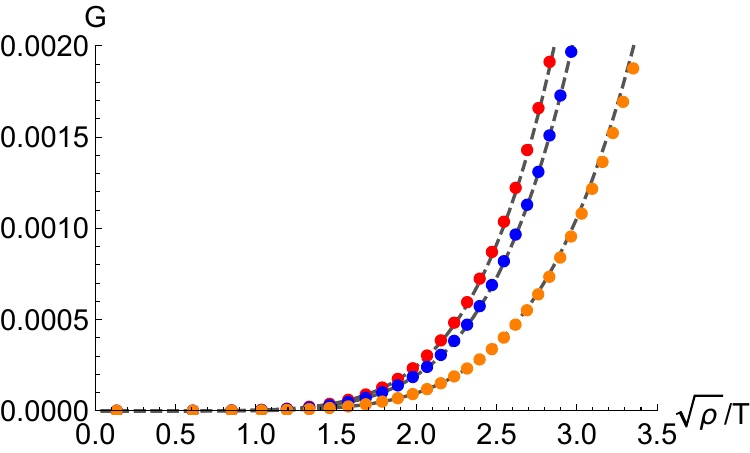}
  \includegraphics[width=0.45\textwidth]{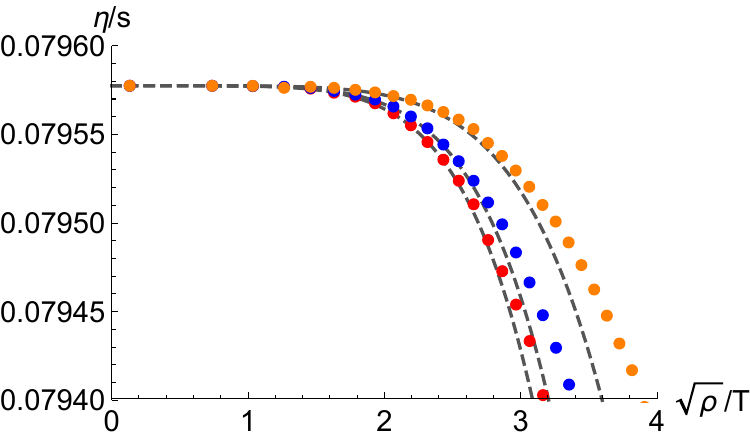}
  \caption{The comparison between the numerical and analytical results of shear elastic modulus and viscosity in the high temperature limit for $\mathcal{K}\in[-1/8, -1/20]$(red-orange). The dots are numerical results obtained by (\ref{Geta}). The gray dashed lines are analytical results obtained by (\ref{AnalyG2}) and (\ref{AnalyEta2}). We have fixed $\lambda=0$ and $\alpha/\sqrt{\rho}=1$.}\label{AnalyModulusEta}
    \vspace{0.3cm}
\end{figure}

The comparison between the numerical and analytical results for shear elastic modulus and viscosity is shown in Fig.\ref{AnalyModulusEta}. 

\subsection{Zero temperature limit}

Next, we will consider $\eta/s$ ratio at zero temperature limit analytically. In this case (\ref{Background}), the black hole becomes extreme and the near-horizon geometry is $AdS_2\times R_2$ \cite{Baggioli:2019rrs}. The equation is given by
\begin{equation}
h''+\left(\frac{f'}{f}-\frac{2}{u}\right)h'+\left(\frac{\omega ^2}{f^2}+\frac{\alpha ^2 \rho ^2 u^4 \mathcal{K}}{2 f \left(\alpha ^2 u^2 \mathcal{K}+1\right)^2}\right)h=0,
\end{equation}
where we have recovered the parameter $\rho$. We follow the method in \cite{Baggioli:2019rrs}. Assume a power-law ansatz for the field $h$
\begin{equation}
h(u)\sim (u-u_h)^\nu.
\end{equation}
The perturbation equation around the near-horizon geometry is
\begin{equation}
\nu \left((\nu-1) u_h+2 u_h\right) f''\left(u_h\right) \left(\alpha ^2 \mathcal{K} u_h^2+1\right){}^2+\alpha ^2 \rho ^2 \mathcal{K} u_h^5=0.
\end{equation}
One can get
\begin{equation}
\nu=-\frac{1}{2}\pm\xi,\qquad \xi \equiv \frac{1}{2}\sqrt{1-\frac{4\alpha ^2 \rho ^2 \mathcal{K} u_h^4}{\left(\alpha ^2 \mathcal{K} u_h^2+1\right)^2 f''\left(u_h\right)}}.\label{xi}
\end{equation}
Thus we can write the general solution 
\begin{equation}
h(u)\sim u^{-1/2-\xi}+\mathcal{G}_{T_{xy}T_{xy}}(\omega)u^{-1/2+\xi}.
\end{equation}

By dimensional analysis, frequency has the dimension of the temperature and the inverse of the radial coordinate ($\omega\sim T \sim u^{-1}$). This implies 
\begin{equation}
\omega^{1/2+\xi}\sim \mathcal{G}_{T_{xy}T_{xy}}(\omega)\,\omega^{1/2-\xi},
\end{equation}
and again
\begin{equation}
\mathcal{G}_{T_{xy}T_{xy}}(\omega)\sim \omega^{2\xi}.
\end{equation}
Therefore
\begin{equation}
\eta\sim T^{2\xi-1}.
\end{equation}
Finally we can obtain
\begin{equation}
\frac{\eta}{s}\sim T^{2\xi-1}.
\end{equation}

Next we analyze $\xi$. In zero temperature limit, one can get
\begin{equation}
f''\left(u_h\right)=-\frac{6}{u_h^2}-\frac{\alpha ^2 \rho ^2 \mathcal{K} u_h^4}{2 \left(\alpha ^2 \mathcal{K} u_h^2+1\right){}^2}+\frac{3 \rho ^2 u_h^2}{2 \left(\alpha ^2 \mathcal{K} u_h^2+1\right)}.
\end{equation}
We set $u_h=1$ and $\alpha/\sqrt{\rho}=1$. Zero temperature limit gives another formula
\begin{equation}
\rho=2 \sqrt{9 \mathcal{K}^2+3}+6 \mathcal{K}.
\end{equation}
Substitute the above relations into (\ref{xi}), we can finally obtain the $\eta/s$ ratio at zero temperature limit
\begin{equation}
\frac{\eta}{s}\sim \left(\frac{T}{\sqrt{\rho}} \right)^{2\xi-1},\qquad \xi=\frac{1}{2} \sqrt{1-\frac{8 \sqrt{3} \mathcal{K}}{\sqrt{3 \mathcal{K}^2+1}}}.\label{xi22}
\end{equation}

\begin{figure}[htbp]
  \vspace{0.3cm}
  \centering
  \includegraphics[width=0.6\textwidth]{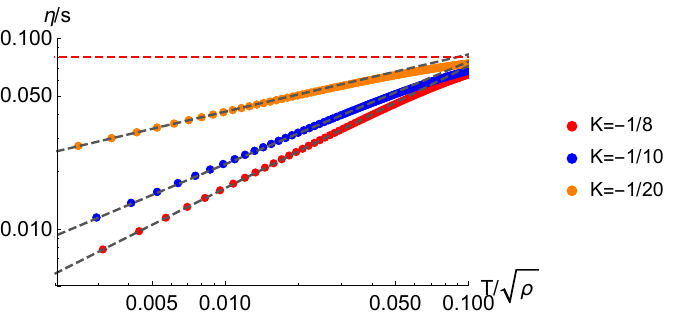}
  \caption{The $\eta/s$ ratio as the function of $T/\sqrt{\rho}$ for $\mathcal{K}\in[-1/8, -1/20]$(red-orange). The dots are numerical results obtained by (\ref{Geta}). The black dashed lines are analytical results obtained by (\ref{xi22}) in the low temperature limit. The red dashed line is the KSS bound. We have fixed coupling $\lambda=0$.}\label{AnalyVis}
    \vspace{0.3cm}
\end{figure}

We show the comparisons of the ratio obtained analytically and numerically in Fig.\ref{AnalyVis}.

\section{Perturbation equations and Green's functions} \label{AppendixEqs}
For convenience in numerical calculations of QNMs, we choose Eddington-Finkelstein (EF) coordinates
\begin{equation}
ds^2=\frac{1}{u^2}\left[-f(u)dt^2-2dtdu+dx^2+dy^2\right].\label{EF}
\end{equation}
Since the boundary field theory is isotropic, we choose the momentum $k$ to be parallel to the $y$-axis. Take the following forms of the transverse perturbations
\begin{equation}
\delta g_{tx}=\frac{1}{u^2}h_{tx}(u)e^{-i\omega t+i k y},\, \delta g_{xy}=\frac{1}{u^2}h_{xy}(u)e^{-i\omega t+i k y},\, \delta A_{x}= a_x(u)e^{-i\omega t+i k y},\, \delta \phi_{x}=\phi_x(u)e^{-i\omega t+i k y},
\end{equation}
and choose radial gauge $\delta g_{ux}=0$ for simplicity.

The equations are
\begin{align}
&0=u (2 i k h_{xy}' (\alpha ^2 u^2 \mathcal{K}+1)^2-2 h_{tx}'' (\alpha ^2 u^2 \mathcal{K}+1)^2+\alpha   \phi_x ' (\rho ^2 u^4 \mathcal{K}-2 \lambda  (\alpha ^2 u^2 \mathcal{K}+1)^2))
\nonumber\\&
+4 h_{tx}' (\alpha ^2 u^2 \mathcal{K}+1)^2+2 \rho  u^3 a_x' (\alpha ^2 u^2 \mathcal{K}+1)^2,\\
&0=2 i \rho  \omega  a_x (\alpha ^2 u^3 \mathcal{K}+u)^2+i (\alpha  \phi_x  \omega  (2 \lambda  (\alpha ^2 u^2 \mathcal{K}+1)^2-\rho ^2 u^4 \mathcal{K})-2 (\alpha ^2 u^2 \mathcal{K}+1)^2 (f k h_{xy}'+\omega  h_{tx}'))
\nonumber\\&
+\alpha  f \phi_x ' (2 \lambda  (\alpha ^2 u^2 \mathcal{K}+1)^2-\rho ^2 u^4 \mathcal{K})+h_{tx} (2 (\alpha ^2 \lambda +k^2) (\alpha ^2 u^2 \mathcal{K}+1)^2-\alpha ^2 \rho ^2 u^4 \mathcal{K})
\nonumber\\&
+2 k \omega  h_{xy} (\alpha ^2 u^2 \mathcal{K}+1)^2,\\
&0=4 i k h_{tx} (\alpha ^2 u^2 \mathcal{K}+1)^2+h_{xy} (\alpha ^2 u (2 \lambda  (\alpha ^2 u^2 \mathcal{K}+1)^2-\rho ^2 u^4 \mathcal{K})+4 i \omega  (\alpha ^2 u^2 \mathcal{K}+1)^2)
\nonumber\\&
-i (\alpha  k u \phi_x  (2 \lambda  (\alpha ^2 u^2 \mathcal{K}+1)^2-\rho ^2 u^4 \mathcal{K})+2 (\alpha ^2 u^2 \mathcal{K}+1)^2 (h_{xy}' (-i u f'+2 i f+2 u \omega )-i f u h_{xy}''+k u h_{tx}')),\\
&0=a_x' (-2 \alpha ^2 f u \mathcal{K}-i (2 \omega -i f') (\alpha ^2 u^2 \mathcal{K}+1))+a_x (k^2 (\alpha ^2 u^2 \mathcal{K}+1)-2 i \alpha ^2 u \omega  \mathcal{K})+\rho  h_{tx}'
\nonumber\\&
-f a_x'' (\alpha ^2 u^2 \mathcal{K}+1),\\
&0=-\phi_x  (k^2 u (\alpha ^2 u^2 \mathcal{K}+1) (2 \lambda  (\alpha ^2 u^2 \mathcal{K}+1)^2-\rho ^2 u^4 \mathcal{K})+2 i \omega  (2 \lambda  (\alpha ^2 u^2 \mathcal{K}+1)^3+\rho ^2 u^4 \mathcal{K} (1-\alpha ^2 u^2 \mathcal{K})))
\nonumber\\&
+\phi_x ' (-2 f (2 \lambda  (\alpha ^2 u^2 \mathcal{K}+1)^3+\rho ^2 u^4 \mathcal{K} (1-\alpha ^2 u^2 \mathcal{K}))+u (f'+2 i \omega ) (\alpha ^2 u^2 \mathcal{K}+1)(2 \lambda  (\alpha ^2 u^2 \mathcal{K}+1)^2-\rho ^2 u^4 \mathcal{K}))
\nonumber\\&
-2 \alpha  h_{tx} (2 \lambda  (\alpha ^2 u^2 \mathcal{K}+1)^3+\rho ^2 u^4 \mathcal{K} (1-\alpha ^2 u^2 \mathcal{K}))-i \alpha  k u h_{xy} (\alpha ^2 u^2 \mathcal{K}+1) (2 \lambda  (\alpha ^2 u^2 \mathcal{K}+1)^2-\rho ^2 u^4 \mathcal{K})
\nonumber\\&
+f u \phi_x '' (\alpha ^2 u^2 \mathcal{K}+1) (2 \lambda  (\alpha ^2 u^2 \mathcal{K}+1)^2-\rho ^2 u^4 \mathcal{K})+\alpha  u h_{tx}' (\alpha ^2 u^2 \mathcal{K}+1) (2 \lambda  (\alpha ^2 u^2 \mathcal{K}+1)^2-\rho ^2 u^4 \mathcal{K}).
\end{align}

The asymptotic behaviour of the bulk fields near the boundary ($u=0$) are
\begin{align}
& h_{tx}=h_{tx(l)}(1+\dots)+h_{tx(s)}u^3(1+\dots),\\
& h_{xy}=h_{xy(l)}(1+\dots)+h_{xy(s)}u^3(1+\dots),\\
& a_{x}=a_{x(l)}(1+\dots)+a_{x(s)}u(1+\dots),
\end{align}
where $l$ and $s$ mark the leading and subleading orders. According to holographic dictionary and standard quantization, we can write the retarded Green's functions as
\begin{align}
&\mathcal{G}^R_{T_{tx}T_{tx}}=\left(2\Delta-d\right)\frac{h_{tx(s)}}{h_{tx(l)}}=3\frac{h_{tx(s)}}{h_{tx(l)}},\\
&\mathcal{G}^R_{T_{xy}T_{xy}}=\left(2\Delta-d\right)\frac{h_{xy(s)}}{h_{xy(l)}}=3\frac{h_{xy(s)}}{h_{xy(l)}},\\
& \mathcal{G}^R_{JJ}=\frac{a_{x(s)}}{a_{x(l)}}.
\end{align}
The physical observables can be extracted from the retarded Green's functions.

\section{DC conductivity and shear viscosity}\label{DCVis}
\begin{figure}[htbp]
  \vspace{0.3cm}
  \centering
  \includegraphics[width=0.48\textwidth]{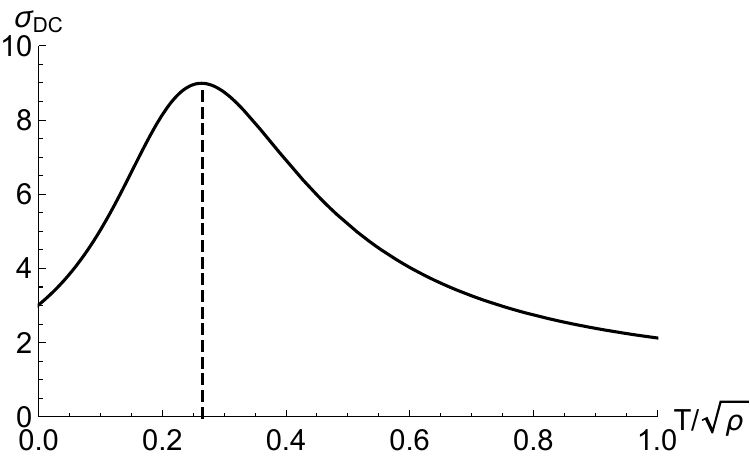}\hspace{5mm}
  \includegraphics[width=0.48\textwidth]{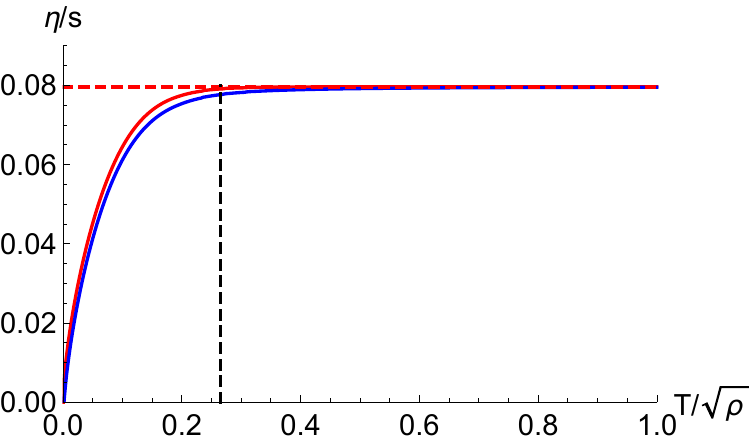}
  \caption{DC conductivity and $\eta/s$ as the function of the temperature. The black dashed line indicates the transition temperature. {\bf Left}: $\mathcal{K}=-1/8$ and $\lambda=5\times 10^{-2}$. {\bf Right}: $\mathcal{K}=-1/8$, $\lambda=0$ (red solid line) and $\lambda=5\times 10^{-2}$ (blue solid line).}\label{figDCVis}
    \vspace{0.3cm}
\end{figure}

We show the figures of DC conductivity and shear viscosity with temperature in Fig.\ref{figDCVis}. As shown in the left figure, there exists the metal-insulator transition near the critical temperature $T/\sqrt{\rho}=0.265$. However there is no specific behaviour for the shear viscosity at both sides of the transition point.

\bibliographystyle{apsrev4-1}
\bibliography{Ref.}
\end{document}